\documentclass[12pt]{iopart}
\usepackage{graphicx}% Include figure files

\usepackage{xcolor}
\usepackage{comment}
\begin{document}

\title{Massless Dirac magnons in the two dimensional van der Waals honeycomb magnet CrCl$_3$}% Force line breaks with \\

%\begin{comment}
\author{Lebing Chen}
\address{Department of Physics and Astronomy, Rice University, Houston, Texas 77005, USA}
\author{Matthew B. Stone, Alexander I. Kolesnikov, Barry Winn}
\address{Neutron Scattering Division, Oak Ridge National Laboratory, Oak Ridge, Tennessee 37831, USA}
\author{Wonhyuk Shon}
\address{Korea Atomic Energy Research Institute, 111 Daedeok-daero, Daejeon 34057, Korea}
\author{Pengcheng Dai}
\ead{pdai@rice.edu}
\address{Department of Physics and Astronomy, Rice University, Houston, Texas 77005, USA}
\author{Jae-Ho Chung}
\ead{jaehc@korea.ac.kr}
\address{Department of Physics, Korea University, Seoul 02841, Korea}
%\end{comment}

\vspace{10pt}
%\begin{indented}
%\item[]August 2017
%\end{indented}

\begin{abstract}
Two dimensional van der Waals ferromagnets with honeycomb structures are expected to host the bosonic version of Dirac particles in their magnon excitation spectra. Using inelastic neutron scattering, we study spin wave excitations in polycrystalline CrCl$_3$, which exhibits ferromagnetic honeycomb layers with antiferromagnetic stackings along the $c$-axis. For comparison, polycrystal samples of CrI$_3$ with different grain sizes are also studied. We find that the powder-averaged spin wave spectrum of CrCl$_3$ at $T$ = 2 K can be adequately explained by the two dimensional spin Hamiltonian including in-plane Heisenberg exchanges only. The observed excitation does not exhibit noticeable broadening in energy, which is in remarkable contrast to the substantial broadening observed in CrI$_3$. Based on these results, we conclude that the ferromagnetic phase of CrCl$_3$ hosts massless Dirac magnons and is thus not topological.
\end{abstract}

\vspace{2pc}
\submitto{\TDM}
\maketitle

\section{Introduction}

Ferromagnets with two-dimensional (2D) van der Waals honeycomb structures are attracting a great deal of research interests in condensed matter physics thanks to their potential for dissipationless spintronic applications \cite{CrX3_Zhang2015,CGT_Gong2017,CrI3_Huang2017,CrBr3_Chen2019,LBChen_PRX2018,LBChen_PRX2021,CrX3_HHKim2019,CrX3_Soriano2020}. In these materials, chemically stable atomic layers are stacked on top of each other via weak van der Waals forces allowing relatively easy exfoliation of single monolayers. The intrinsic 2D ferromagnetism was first observed in Cr$_2$Ge$_2$Te$_6$ and CrI$_3$ where the Curie temperatures remain finite down to their monolayer limits \cite{CGT_Gong2017,CrI3_Huang2017}. These experimental results demonstrate that the magnetic anisotropy in these 2D materials are strong enough to overcome thermal fluctuations inherent in low-dimensional magnets \cite{CrI3_Lado2017,CrI3_DHKim2019}. In CrI$_3$, the interlayer exchange couplings between ferromagnetic layers were observed to change from ferromagnetic to antiferromagnetic as the number of stacked monolayers decreased \cite{CrI3_Jiang2019,interlayer_SWJang2019,CrI3_Song2019_Pressure,CrI3_Li2019_Pressure}. Such remarkable behavior is ascribed to the changes in the mode of stacking, for instance, from rhombohedral to monoclinic. Similar stacking-dependent 2D orderings were soon observed also in isostructural CrBr$_3$ and CrCl$_3$ \cite{CrBr3_Chen2019,CrBr3_Zhang2019,CrCl3_Klein2019}. 

While the above discoveries raise hopes for 2D spin devices for spintronics \cite{CrI3_Zhong2017,CrBr3_Ghazaryan2018}, to realize such practical applications it is important to understand their thermal properties in detail. Thermal excitations in long-range ferromagnetic ordered materials appear as spin waves, or magnons as their quantized units. Given the structural similarity, the magnon band structures of honeycomb ferromagnets are expected to look analogous to the electronic band structure of graphene \cite{DiracMaterials2014,DiracMagnon2018}. The latter is well known to exhibit linear energy-momentum relations ($E\propto p$) near its Fermi energy, where conduction electrons behave as relativistic massless Dirac particles overcoming wave-particle duality \cite{graphene}. It was thus predicted that the bosonic version of Dirac particles, or Dirac magnons, should also appear on the magnon bands of honeycomb ferromagnets \cite{DiracMagnon2018}. Such linear magnon dispersion relations have experimentally been observed earlier in CrBr$_3$ \cite{CrBr3_Samuelson1971}, and more recently in CoTiO$_3$ \cite{CoTiO3_Yuan2020,CoTiO3_Elliot2021} and a three dimensional antiferromagnet Cu$_3$TeO$_6$ \cite{Cu3TeO6_Yao2018,Cu3TeO6_Bao2018}. In the spintronic transport, one of the key research interests lies in whether or not these Dirac magnons will be massive and thereby open an insulating gap in the bulk bands. If the bulk gap opens by breaking the time-reversal symmetry at the Dirac point, the accompanying edge states will be gapless allowing topological transport \cite{Haldane1988,kanemele2005}. One of the best examples of such topological band structure is found in 2D graphene, where spin-orbit coupling opens a small electronic band gap ($\sim$ 1 $\mu$eV) at its six-fold Dirac wave vectors, $Q =(\frac{1}{3}, \frac{1}{3})$. In the case of honeycomb ferromagnets, the antisymmetric Dzyaloshinkii-Moriya (DM) interactions between the next nearest neighbor spins may open topological energy gaps at the magnon Dirac points \cite{Owerre2016}. The resultant magnon band structure will be separated into two, one higher and the other lower in energy, in which Berry curvatures are integrated to the integer Chern numbers $C^{\pm}=\pm1$, respectively \cite{SKKim2016}. Recently, finite energy gaps were indeed observed in the magnon bands of CrI$_3$ \cite{LBChen_PRX2018,LBChen_PRX2021,LBChen_PRB2020} and CrBr$_3$ \cite{CrBr3_Cai2021}, respectively, indicating that the Dirac magnons in these ferromagnets are massive and topological.

\begin{figure}[t]
\centering
\includegraphics[scale=0.60]{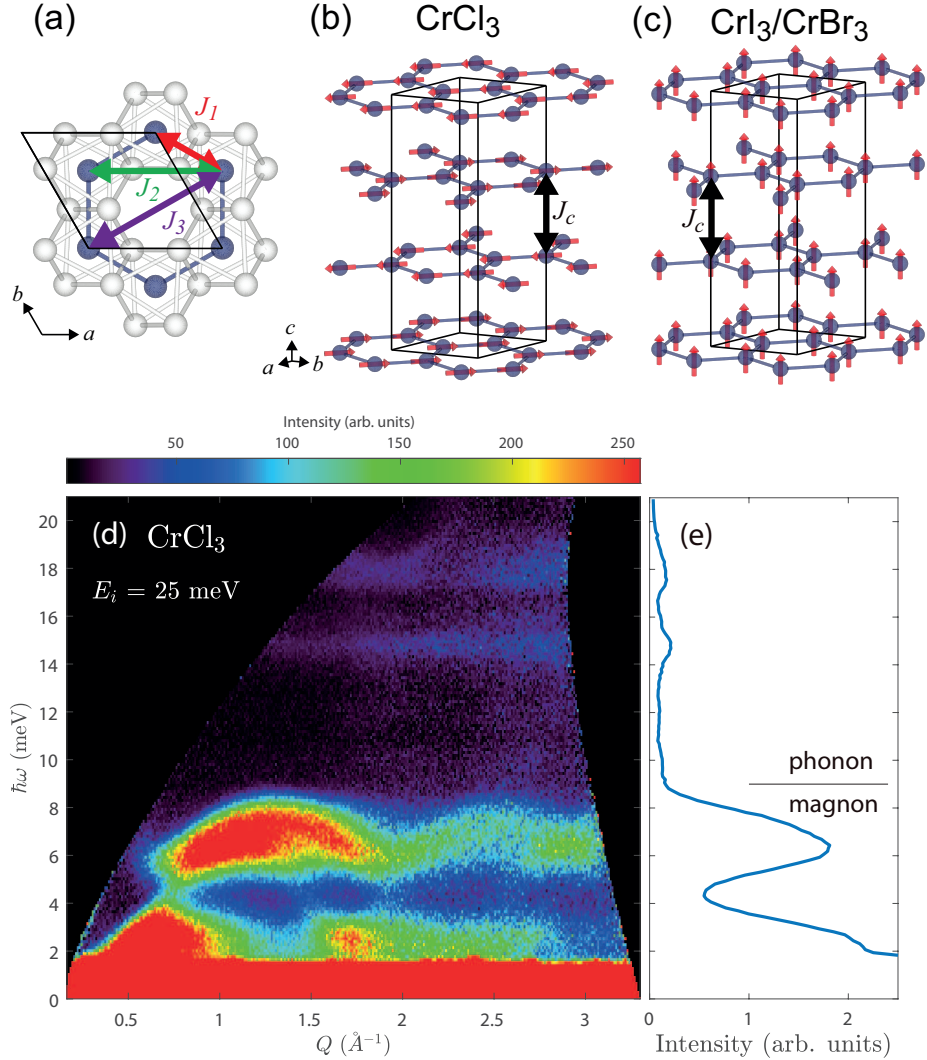}
\caption{\label{crcl3_sqw}(a) The honeycomb structure of a van der Waals layer in CrX$_3$ (X = Cr, I) (b,c) Spin ordering and stacking patterns of (b) antiferromagnetic CrCl$_3$ and (c) ferromagnetic CrI$_3$. In (b,c), only Cr ions are shown for simplicity. (d) Inelastic neutron scattering intensity of CrCl$_3$ at $T$ = 4 K measured with $E_i$ = 25 meV. The intensity integrated over the entire momentum transfer is plotted in (e).
}
\end{figure}

CrCl$_3$ is unique among these van der Waals honeycomb ferromagnets because in the magnetically ordered phase its spins lie within the honeycomb planes and the adjacent monolayers are antiferromagnetically coupled (see Figs. \ref{crcl3_sqw}(a) and \ref{crcl3_sqw}(b)) \cite{CrCl3_McGuire2017,CrCl3_Dupont2021}. These properties are in contrast to CrI$_3$, CrBr$_3$ and Cr$_2$Ge$_2$Te$_6$ where spins are oriented perpendicular to the honeycomb planes and coupled ferromagnetically between monolayers (see Fig. \ref{crcl3_sqw}(c)). Since its spins are lying in the plane, CrCl$_3$ is expected to host massless Dirac magnons without gap openings at the Dirac points. The apparent easy-plane anisotropy in CrCl$_3$ is identified as the magnetic shape anisotropy of the dipole-dipole interaction that barely overcomes the weak spin-orbit coupling of light Cl$^-$ ligand ions \cite{CrCl3_Lu2020}. Whereas the interlayer exchanges may greatly enhance as the stacking changes down to bilayers \cite{CrCl3_Klein2019,CrCl3_Seri2020}, the bulk bands in CrCl$_3$ will exhibit nearly 2D magnons owing to its remarkably weak interlayer couplings. 
 
In this work using inelastic neutron scattering, we report the powder-averaged spin wave excitation spectrum of CrCl$_3$ and show that its magnon at the Dirac wave vector is indeed gapless and well-described by the 2D spin Hamiltonian including isotropic Heisenberg exchanges only. By comparing the results with the powder-averaged spectrum of CrI$_3$, we also find that the magnons in CrCl$_3$ do not exhibit broadening in energy, which is in remarkable contrast to other van der Waals honeycomb ferromagnets. 

\section{Methods}

The samples used in this study were prepared in a few different forms. The CrCl$_3$ sample used for inelastic neutron scattering measurements was commercially purchased from Alfa Aesar in polycrystalline form with 99.9 \% purity. For dc magnetization measurements, the commercial CrCl$_3$ was recrystallized into relatively larger platelets using the chemical vapor transport method \cite{CrCl3_McGuire2017}. The CrI$_3$ samples were initially obtained from batches of crystal growth, for which we used the chemical vapor transport method using I$_2$ as the transport agent \cite{CrI3_McGuire2014}. Two CrI$_3$ samples were then prepared with relatively large (L) or small (S) grain sizes, respectively. The sample L consisted of small crystallites of approximately 1 mm in width which were collected as-is directly from the growth batch without further treatment. In contrast, the sample S was pulverized by milling with alumina balls overnight. The typical size of the pulverized grains was at sub-micrometer level when viewed under an optical microscope. Both types of CrI$_3$ samples were used for inelastic neutron scattering and dc magnetization measurements. In the Supplemental Information, we show the comparison of the elastic parts of the neutron scattering data between the two CrI$_3$ samples confirming that the Bragg peaks in the sample S are significantly broader that those in the sample L \cite{SI}. We also observe a consistent difference in the x-ray diffraction data collected at room temperature \cite{SI}. 

Our inelastic neutron scattering experiments were performed using two time-of-flight spectrometers at the Spallation Neutron Source, Oak Ridge National Laboratory. Spin wave excitations of CrCl$_3$ were measured on the SEQUOIA spectrometer using incident neutron energies of $E_i$ = 25, 12, or 4 meV \cite{SEQUOIA}. We used $\sim$ 1g sample containing a large number of randomly oriented polycrystalline pieces of CrCl$_3$, which were sealed in an aluminum can with helium for thermal exchange. The excitations of the two CrI$_3$ samples were measured on the HYSPEC spectrometer using an incident neutron energy of $E_i$ = 25 meV \cite{HYSPEC}. The sample mass is $\sim$ 6 g for both L and S samples, respectively.

To analyze the inelastic neutrons scattering intensities, we calculated the neutron scattering cross section of the magnon excitations using the SpinW program \cite{SpinW}. The magnon calculation was based on the linear spin wave approximation within the Holstein-Primakoff formalism. The least square fitting was performed between the observed and calculated intensities over the same wave vector ranges, from which we obtained the best-fit exchange constants. The model spin Hamiltonian and the parameters included will be discussed in the next section.

\section{Results and discussions}

\subsection{DC magnetization}

\begin{figure}[t]
\centering
\includegraphics[scale=0.60]{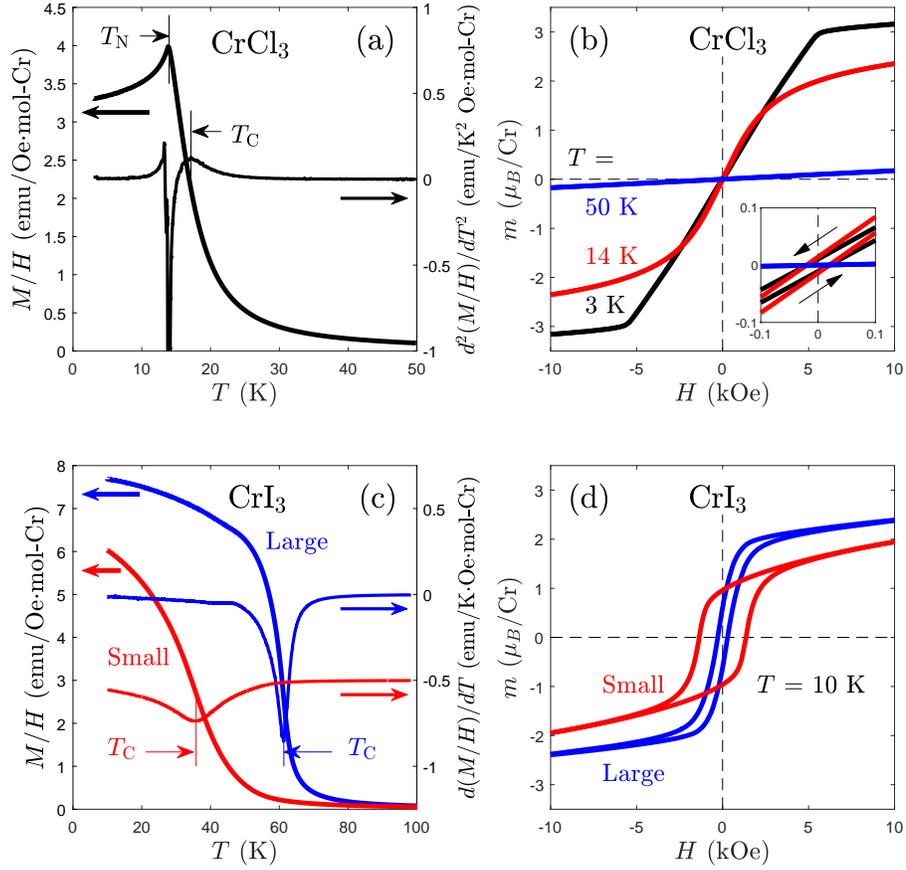}
\caption{\label{magnetization} (a) Temperature dependence of the dc magnetization, $M(T)/H$, and its second derivative, $d^2(M(T)/H)/dT^2$ (thin lines and right axis), of CrCl$_3$ with an external field $H$ = 1 kOe applied along the $c$ axis. (b) The magnetic hysteresis curves of CrCl$_3$ at selected temperatures with external magnetic field along the $c$ axis. The inset shows the data around zero field. (c) Temperature dependencies of the dc magnetization, $M(T)/H$, and their derivatives, $d(M(T)/H)dT$ (thin lines and right axis), of the large and small powder samples of CrI$_3$, respectively. The derivative for the small sample is vertically shifted for clarity. (d) The magnetic hysteresis curves of the large and small powder samples of CrI$_3$ at $T$ = 10 K.
}
\end{figure}

We first discuss the dc magnetization data of CrCl$_3$ and two different CrI$_3$ crystalline samples, which are summarized in Fig. \ref{magnetization}. The measurements were done with dc magnetic field applied along the $c$ axis taking advantage of the distinctive platelike shape of the van der Waals crystals. We note, however, that the orientations of the CrI$_3$-S platelets were relatively uncertain. 

In CrCl$_3$, the antiferromagnetic transition is easily noticed as a sharp kink at $T_\mathrm{N}$ = 14.0 K in the temperature dependent magnetization ($M/H$) plotted in Fig. \ref{magnetization}(a). Since its inverse magnetization deviated significantly from linearity, the ferromagnetic transition could not be identified by fitting with the Curie-Weiss law. The Curie temperature of CrCl$_3$ has previously been identified at $T_\mathrm{C}$ = 17.2 K by heat capacity measurement \cite{CrCl3_McGuire2017}. In our current dc magnetization data, we find that the maximum of its second derivative, $d^2(M/H)/dT^2$, is located at the same temperature [see Fig. \ref{magnetization}(a), right axis]. We consider it to be the better estimate of $T_\mathrm{C}$ than the maximum of the first derivative, which is located at a lower temperature by $\sim$ 2 K \cite{CrCl3_Seri2020}. 

The field dependence of the magnetization of CrCl$_3$ is plotted in Fig. \ref{magnetization}(b) as $m$-$H$ hysteresis curves, where $m$ is the magnetic moment per Cr ion. Three temperatures are selected to represent the paramagnetic ($T$ = 50 K), ferromagnetic ($T$ = 14 K), and antiferromagnetic ($T$ = 3 K) phases, respectively. Whereas the magnetic moment in the paramagnetic phase changes weakly over the entire range below 10 kOe, in the antiferromagnetic phase exhibits a sharp change in the slope at around 5.6 kOe followed by gradual saturation to the value of $m \approx 3\mu_B$. In the intermediate and ferromagnetic phase, the magnetic moment does not linearly depend on the magnetic field. The inset in Fig. \ref{magnetization}(b) shows there are small but finite remnant magnetizations, $\sim 0.014\mu_B$, which are nearly identical between the two phases. It thus suggests that the ferromagnetic orderings within van der Waals monolayers remain unchanged below $T_\mathrm{C}$. 

The temperature dependences of dc magnetization of the two CrI$_3$ samples, L and S, are plotted in Fig. \ref{magnetization}(c). While both samples show expected transitions from paramagnetic to ferromagnetic, $T_\mathrm{C}$ is noticeably smaller in the sample S which was pulverized with ball milling. Contrary to the case of CrCl$_3$, in CrI$_3$-L we find the maximum of its first derivative is located closer to the previously known Curie temperature at $T_\mathrm{C}$ = 61.3 K than the second derivative. We thus use the same method to identify the Curie temperature of CrCl$_3$-S at $T_\mathrm{C}$ = 35.8 K. Note that this is quite close to the Curie temperature of CrI$_3$ in the monolayer limits \cite{CrI3_Huang2017}. Therefore, we conclude that the effect of ball-milling was in significantly reducing not only the area of van der Waals layers but also the number of stacked layers. 

Another evidence for effective thinning of the sample S is found in the enhanced coercivity revealed in the $m-H$ hysteresis curves. As plotted in Fig. \ref{magnetization}(d), the sample S has significantly larger coercive field ($H^S_c \approx$ 1.3 kOe) than the sample L at 0.30 kOe although the former has smaller moment in zero field. This behavior is consistent not only with the previous independent observations of the bulk and monolayer samples \cite{CrI3_Huang2017,CrI3_McGuire2014}, but also the recent report of surface layers being harder than the bulk by the magnetic force microscopy measurements \cite{CrI3_Niu2020}. Therefore, we conclude that the sample S is closer to the monolayer limits than the sample L.

\subsection{Spin wave excitations in CrCl$_3$}

Overview of the inelastic neutron scattering intensities of CrCl$_3$ at $T$ = 4 K is displayed in Fig. \ref{crcl3_sqw}(d). Strong and dispersive excitations are observed at low energies below $E \le$ 9 meV, whose intensity decreases at larger wave vectors ($Q$) indicating that these are magnetic excitations. These excitations consist of two bands separated at $\sim$ 4 meV, which are ascribed to the two sublattice spins of a simple honeycomb lattice. The upper and lower bands account for in-phase and out-of-phase precessions, respectively, of the two sublattice spins. Additional intensities observed at high energies, $E \approx$ 14.5 and 17.5 meV, are ascribed to phonon excitations as they increase in intensity at higher wave vectors. These energies are also consistent with recent Raman scattering observations of phonon signals at $\sim$ 115 and 140 cm$^{-1}$ \cite{CrX3_Zhang2015,CrCl3_Klein2019}. 

\begin{figure}[t]
\centering
\includegraphics[scale=0.60]{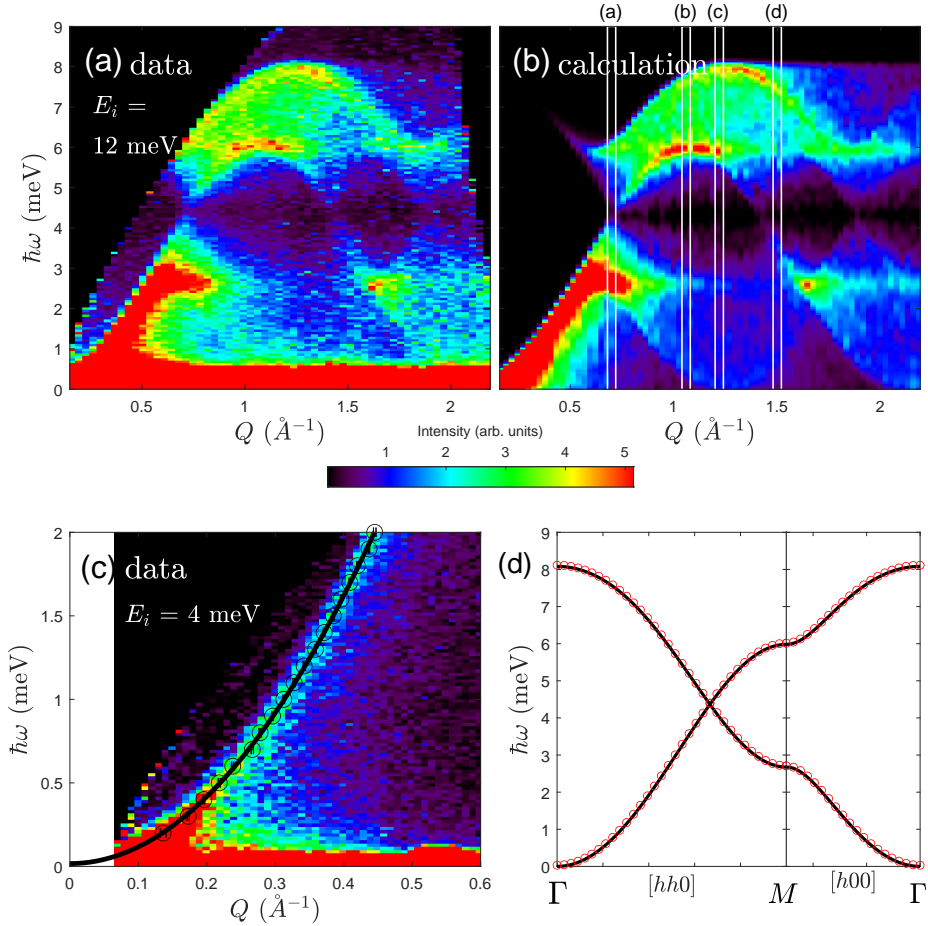}
\caption{\label{crcl3_spinwaves} (a) The powder-averaged spin wave excitation intensities of CrCl$_3$ measured at $T$ = 4 K with $E_i$ = 12 meV and (b) the calculated intensities convoluted with the instrumental resolution. The best-fit values of $J_1$, $J_2$ and $J_3$ used in (b) are listed in the main text. The double vertical lines and the alphabets above them mark the ranges of integration for constant-$Q$ plots in Fig. \ref{crcl3_constQ} and the corresponding subplot labels, respectively. (c) The low-energy part of the spin wave excitations measured with $E_i$ = 4 meV. The black solid line is the quadratic fit to the peak positions marked with open circles. (d) Calculated spin wave energies along selected high-symmetry directions. The solid lines are the calculations using $J_1$, $J_2$ and $J_3$ only while the red circles also include $J_c$ = 0.0077 meV and $D$ = 0.014 meV.
}
\end{figure}

To observe the details of spin wave excitations, we repeated the measurements with instrumental setups with higher energy resolutions. First, the data obtained with the incident energy at $E_i$ = 12 meV is shown in Fig. \ref{crcl3_spinwaves}(a). The upper band includes two distinctive intensity maxima due to powder averaging: one near $E \approx$ 6 meV and the other 8 meV. The maximum at 6 meV corresponds to the van Hove singularity at $Q_M$ = ($\frac{1}{2}$, $\frac{1}{2}$, 0) while the other at 8 meV corresponds to the top of the magnon excitation at $Q_\Gamma$ = (1, 0, 0). These two wave vectors are located at the boundary and the center, respectively, of the 2D Brillouin zone of the honeycomb lattice. The overall band structure is consistent with the nearest neighbor exchange ($J_1$) being ferromagnetic just as in CrBr$_3$ and CrI$_3$. Given the antiferromagnetic coupling between the adjacent ferromagnetic layers and the number of layers per unit cell being three, the dispersion minimum along the $c$ axis is expected to appear at $Q_Z$ = (0 0 $\frac{3}{2}$), or $|Q_Z|$ = 0.54 \AA$^{-1}$. At the same time, the zero wave vector at $Q$ = 0 will correspond to the maximum of the dispersion along the $c$ axis which is determined by the strengths of $J_c$ and the magnetic anisotropy. The plots in Fig. \ref{crcl3_spinwaves}(c) shows the low-energy magnon mode around $Q$ = 0 measured with the incident energy at $E_i$ = 4 meV with the full-width-half-maximum energy resolution of 0.09 meV at the elastic channel. We attempted to estimate the dispersion maximum at $Q$ = 0 by fitting these magnon energies with a typical quadratic function of momentum, $\hbar\omega = aQ^2 + \Delta$. As plotted with a solid line in Fig. \ref{crcl3_spinwaves}(c), the intercept is found at $\Delta$ = 0.016 meV which is several times smaller than the instrumental resolution. We thus tentatively conclude that the strengths of $J_c$ and the magnetic anisotropy are too small to be resolved by the current experimental configurations.

To fit the spin wave spectrum and extract the exchange constants, we consider the spin Hamiltonian including three intra-plane ($J_1$, $J_2$, $J_3$) and one inter-plane ($J_c$) isotropic Heisenberg exchanges depicted in Fig. \ref{crcl3_sqw}(a) as well as the easy-plane anisotropy ($D>0$):

\begin{equation}
H = \frac{1}{2}\sum_{ij}J_{ij}\mathbf{S}_i\cdot\mathbf{S}_j + \sum_jD(\mathbf{S}_j\cdot\hat{c})^2.
\end{equation}

\noindent Since the observed spectra do not reveal low-lying branches perpendicular to the honeycomb planes, we attempted to estimate the values of $J_c$ and $D$ from the existing magnetization data in the literature. The dc magnetization of bulk CrCl$_3$ is reported to saturate at $B^{sat}_{ab}$ = 0.20 T or $B^{sat}_c$ = 0.56 T along the direction parallel or perpendicular to the honeycomb planes, respectively \cite{CrCl3_Seri2020,CrCl3_MacNeill2019}. Given that antiferromagnetic $J_c$ must be overcome to arrive at the in-plane saturation, we obtain $J_c = g\mu_B B^{sat}_{ab}/2S$ = 0.0077 meV where we used $g$ = 2 and $S$ = 3/2. As both $J_c$ and $D$ must be simultaneously overcome for the out-of-plane saturation, we may subsequently obtain $D = g\mu_B B^{sat}_{c}/S - J_c$ = 0.014 meV. We find that the magnon bandwidth along the [0 0 $L$] direction becomes as small as 0.03 meV when these two parameters are included. The in-plane dispersions with and without them, respectively, are plotted in Fig. \ref{crcl3_spinwaves}(d). For this reason, we included only the three in-plane Heisenberg exchanges to perform least square fittings of the linear spin wave intensities against the inelastic neutron scattering data. The resultant best fit parameters are $J_1 = -0.95$ meV, $J_2 = -0.024$ meV, and $J_3$ = 0.051 meV, and the calculations using these parameters are shown in Fig. \ref{crcl3_spinwaves}(b). The values of these parameters are also listed in Table \ref{Js} along with those previously reported for CrBr$_3$ and CrI$_3$.

\begin{figure}[t]
\centering
\includegraphics[scale=0.60]{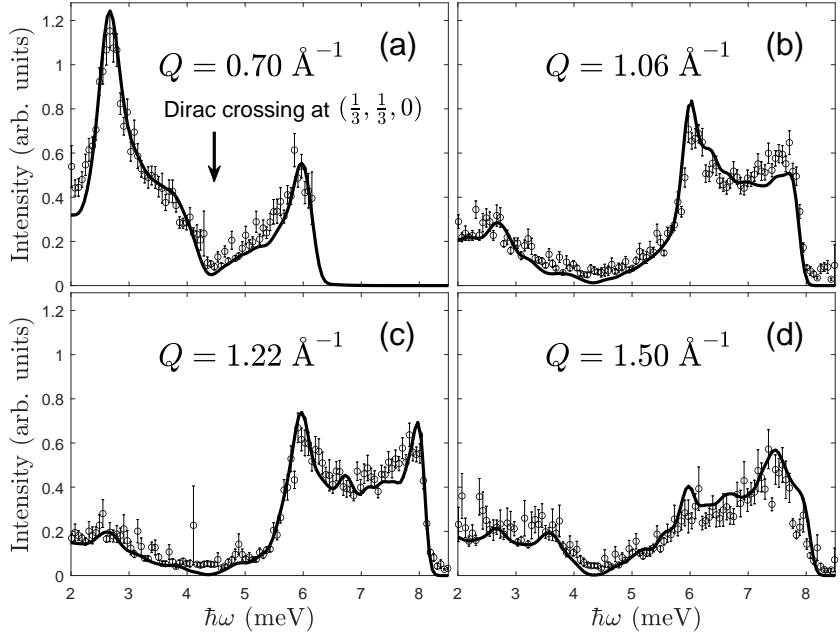}
\caption{\label{crcl3_constQ} Energy dependencies of the spin wave excitation intensities at momentum transfer values ($Q$). The ranges of integration are marked as double vertical lines in Fig. \ref{crcl3_spinwaves}(b). The open circles are the inelastic neutron scattering data measured with $E_i$ = 12 meV, which are integrated over $\pm$ 0.02 \AA$^{-1}$. The solid lines are the calculated intensities using the parameters discussed in the text convoluted with the energy-dependent instrumental resolutions.
}
\end{figure}

Since large topological energy gaps have been observed in the magnon bands of CrI$_3$ and CrBr$_3$ \cite{LBChen_PRX2018, LBChen_PRX2021, CrBr3_Cai2021}, it is important to check whether or not such gaps appear also at the Dirac points of CrCl$_3$. Although the excitation spectrum in Fig. \ref{crcl3_spinwaves}(a) is powder-averaged, its Dirac point at $|\mathbf{Q}_{\frac{1}{3}\frac{1}{3}0}|$ = 0.70 \AA$^{-1}$ is clearly visible at $\hbar\omega$ = 4.4 meV. The apparent intensity discontinuity between the upper and lower modes is observed at the wave vector corresponding to the Dirac crossings of the linear dispersions. To check for the possible of a gap opening, we plot a few selected constant-$Q$ cuts of the inelastic neutron scattering intensities including the Dirac wave vector in Fig. \ref{crcl3_constQ} . The calculated intensities, which are convoluted by the instrumental resolutions only, are plotted together as solid lines. The excellent agreement between the observations and calculations indicates that there is no additional broadening in the excitation. Such behavior is in remarkable contrast to the excitation spectra of CrI$_3$ or CrBr$_3$, which exhibit significant broadening in energy throughout their magnon bands at low temperatures\cite{LBChen_PRX2018, LBChen_PRX2021, CrBr3_Cai2021}. The magnon broadening in insulating ferromagnets typically occurs via magnon-magnon interactions or magnon-phonon coupling as the temperature approaches $T_\mathrm{C}$ \cite{mmi1,mmi2,mpc1}. We thus stress that the resolution-limited magnon in CrCl$_3$ at $T =$ 4 K is considered normal whereas the magnon broadening observed in CrI$_3$ and CrBr$_3$ are anomalous. Since the magnon broadening in these van der Waals ferromagnets is beyond the scope of the current work, this subject will be investigated in a separate work. Taking this into account, we find that the constant energy cut at $|\mathbf{Q}_{\frac{1}{3}\frac{1}{3}0}|$ = 0.70 \AA$^{-1}$ is well explained by a calculation without a gap opening contrary to CrI$_3$ and CrBr$_3$. The absence of the Dirac gap in CrCl$_3$ is consistent with its spins being oriented parallel to the honeycomb plane in zero field. Note that the gap openings in CrI$_3$ and CrBr$_3$ are ascribed to the Dzyaloshinkii-Moriya vectors oriented perpendicular to the honeycomb planes and parallel to the ordered spins \cite{LBChen_PRX2018, CrBr3_Cai2021}.

\begin{table}[]
\centering
\begin{tabular}{cccccccccc}
\hline
      & $T_\mathrm{N}$(K) & $T_\mathrm{C}$(K) & $J_1$ & $J_2$ & $J_3$ & $J_c$ & $DM$ & $D$ & note      \\ \hline
CrCl$_3$ & 14.0 & 17.2 &  -0.95  & -0.024  & 0.051 &  0.008$^\dag$ & - & 0.014$^\dag$  & This work \\
CrBr$_3$ & - & 32 &  -1.36  & -0.06 & 0.12 & -  & 0.07  & -0.04 & Ref. \cite{CrBr3_Cai2021}     \\
CrI$_3$  & - & 61 &  -2.01 &  -0.16 &  0.08 & -0.59   & 0.31  & -0.22   & Ref. \cite{LBChen_PRX2018}      \\ \hline
\end{tabular}
\caption{\label{Js}Comparison of exchange parameters among CrCl$_3$, CrBr$_3$ and CrI$_3$. $DM$ is the strength of DM exchange. All units are in meV except the temperatures in K. Note that, for CrI$_3$, an updated set of parameters are reported in Ref. \cite{LBChen_PRX2021} including additional interplane exchanges that are not shown in Figs. \ref{crcl3_sqw}(b) or \ref{crcl3_sqw}(c). ($^\dag$: Not used for the spin wave calculations.)}
\end{table}

\subsection{Spin wave excitations in CrI$_3$}

\begin{figure}[t]
\centering
\includegraphics[scale=0.60]{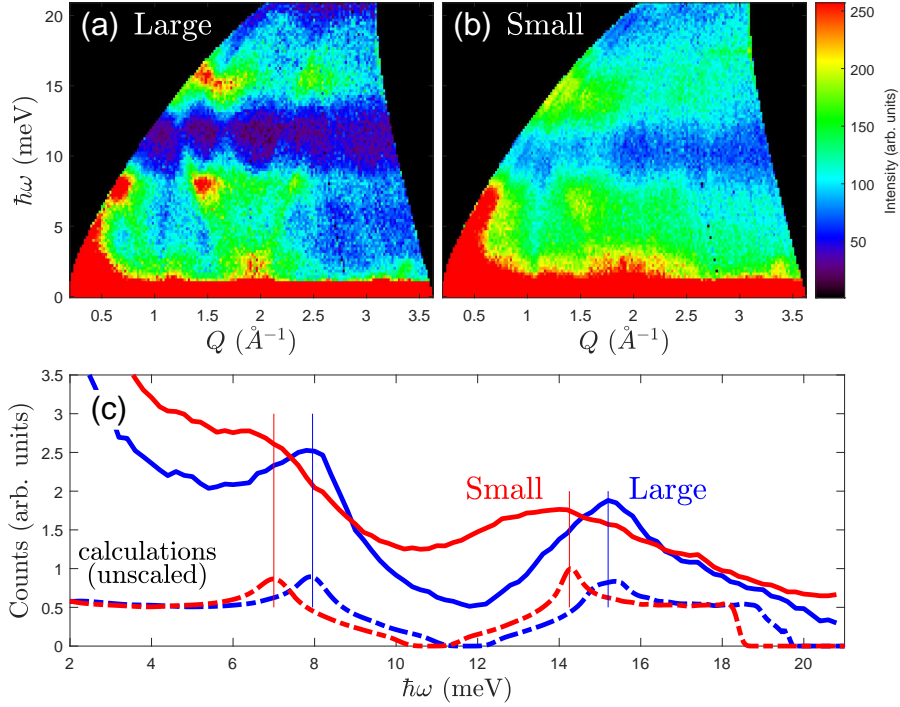}
\caption{\label{cri3_spinwaves} (a,b) The powder-averaged inelastic neutron scattering intensities of (a) L and (b) S samples of CrI$_3$ measured at $T$ = 2 K with $E_i$ = 25 meV. (c) Energy dependences of the $Q$-integrated inelastic neutron scattering intensities. Solid lines are the experimental data while dashed lines are the calculated intensities using the parameters reported in Ref. \cite{LBChen_PRX2021}. Note that the latter was convoluted with the energy-dependent instrumental resolutions. 
}
\end{figure}

We have just shown in CrCl$_3$ that the ratio of its net inter-planar to net intra-planar exchanges per Cr$^{3+}$ spin is $3\times 10^{-3}$. This ratio is a few orders of magnitude smaller than $\sim 9\times 10^{-1}$ in CrI$_3$ \cite{LBChen_PRX2018, LBChen_PRX2021}. Since spin wave energies of CrCl$_3$ will hardly change upon inclusions of the inter-planar exchanges, the excitations observed in this work must be virtually identical to the magnons in the monolayer limit. In the case of CrI$_3$, 2D magnon energies have been observed by means of Raman scattering in the monolayer limit \cite{CrI3_Cenker2021}. These energies are largely consistent with those of the bulk crystals when the interplanar exchanges are excluded. To observe magnon energies in CrI$_3$ close to the 2D limit, we performed inelastic neutron scattering measurements on the two polycrystalline samples with different amounts of thinning. As previously described, the sample S is mechanically reduced down close to the 2D limit. The momentum-integrated neutron scattering intensities of the sample L and S, respectively, at $T$ = 2 K are plotted in Fig. \ref{cri3_spinwaves}. Commonly in both samples, the powder-averaged excitation spectra exhibit two distinct bands corresponding to the upper and lower magnon modes, respectively, expected for honeycomb ferromagnets. Their overall band structures are equivalent to that of CrCl$_3$ (see Fig. \ref{crcl3_spinwaves}) although the excitations in the sample S are significantly broader in energy. It thus suggests that CrI$_3$ still retains its in-plane honeycomb structure after the ball milling.

When we directly compare the integrated magnon densities of states, it becomes clear that the two intensity peaks of the sample S are shifted to lower energy by $\sim$ 1 meV. To understand the origin of the observed energy shift, we performed the linear spin wave calculations based on the spin Hamiltonian derived recently from the single crystal excitations. In Fig. \ref{cri3_spinwaves}(c), the blue dash-dotted line is calculated using the exact three dimensional spin Hamiltonian reported in \cite{LBChen_PRX2021} while the red dash-dotted line is the 2D calculation excluding the inter-planar exchanges. We note that both lines were convoluted with the energy-dependent instrumental resolutions. Remarkably, the two maxima of the 2D calculation excellently match those of the sample S. Such agreement strongly suggests that the ordered stacking of the van der Waals layers in this sample is effectively compromised resulting in nearly 2D magnetism. It also confirms that the bulk excitations of CrI$_3$ are far from 2D magnons. We also note that spectra of both samples are significantly broader than the instrumental resolutions. Such behavior is consistent with the broadening already observed in the bulk crystal excitations, and is thus most likely intrinsic to the CrI$_3$. 

\section{Conclusions}

In this work, we used inelastic neutron scattering to study powder-averaged magnon excitations in the van der Waals honeycomb ferromagnets CrCl$_3$ and CrI$_3$. While their magnon band structures were qualitatively equivalent, we found that magnons in CrCl$_3$ did not exhibit broadening in energy contrary to CrI$_3$ where significant broadening was observed. The magnon excitation spectrum of CrCl$_3$ could be reproduced using the 2D spin Hamiltonian including in-plane Heisenberg exchange only and without considering the energy gap opening at the Dirac point, which are also in contrast to CrI$_3$. Based on these observations, we conclude that the 2D Dirac magnon in CrCl$_3$ is massless and not topological.

\section{Acknowledgments}
%For the sake of anonymising, the acknowledgements will be added later.
The authors are grateful to Long Qian, Yuxiang Gao, and Emilia Morosan for their assistance in XRD measurements. This work is supported by he National Research Foundation (NRF) of Korea (Grant nos. 2020R1A5A1016518 and 2020K1A3A7A09077712). The neutron scattering and sample growth work at Rice is supported by U.S. DOE BES under Grant No. DE-SC0012311 and by the Robert A. Welch Foundation under Grant No. C-1839, respectively (P.D.). WS was supported by the internal R\&D programme at KAERI (Grant No. 524460-21). A portion of this research used resources at the Spallation Neutron Source, a DOE Office of Science User Facilities operated by the Oak Ridge National Laboratory.

{}


\section{References}

\begin{thebibliography}{}

\bibitem{CrX3_Zhang2015} Zhang W-B, Qu Q, Zhua P and Lam C-H 2015 Robust intrinsic ferromagnetism and half semiconductivity in stable two-dimensional single-layer chromium trihalides \textit{J. Mater. Chem.} C \textbf{3} 12457-68

\bibitem{CGT_Gong2017} Gong C \textit{et al} 2017 Discovery of intrinsic ferromagnetism in two-dimensional van der Waals crystals \textit{Nature} \textbf{546} 265-9

\bibitem{CrI3_Huang2017} Huang C \textit{et al} 2017 Layer-dependent ferromagnetism in a van der Waals crystal down to the monolayer limit \textit{Nature} \textbf{546} 270-3

\bibitem{CrBr3_Chen2019} Chen W \textit{et al} 2019 Direct observation of van der Waals stacking–dependent interlayer magnetism \textit{Science} \textbf{366} 983–7

\bibitem{LBChen_PRX2018} Chen L \textit{et al} 2018 Topological Spin Excitations in Honeycomb Ferromagnet CrI$_3$, \textit{Phys. Rev. X} {\bf 8} 041028

\bibitem{LBChen_PRX2021} Chen L \textit{et al} 2021 Magnetic field effect on topological spin excitations in CrI$_3$, \textit{Phys. Rev. X} {\bf 11} 031047

\bibitem{CrX3_HHKim2019} Kim H H \textit{et al} 2019 Evolution of interlayer and intralayer magnetism in three atomically thin chromium trihalides \textit{Proc. Nat. Acad. Sci.} \textbf{116} 11131-6

\bibitem{CrX3_Soriano2020} Soriano D, Katsnelson M I and Fernandez-Rossier J 2020 Magnetic Two-Dimensional Chromium Trihalides: A Theoretical Perspective \textit{Nano Lett.} \textbf{20} 6225-34

\bibitem{CrI3_Lado2017} Lado J L and Fernández-Rossier J 2017 On the origin of magnetic anisotropy in two dimensional CrI$_3$ \textit{2D Mater.} \textbf{4} 035002

\bibitem{CrI3_DHKim2019} Kim D-H \textit{et al} 2019 Giant Magnetic Anisotropy Induced by Ligand LS Coupling in Layered Cr Compounds \textit{Phys. Rev. Lett.} \textbf{122} 207201

\bibitem{CrI3_Jiang2019} Jiang J \textit{et al} 2019 Stacking tunable interlayer magnetism in bilayer CrI$_3$ \textit{Phys. Rev. B} \textbf{99} 144401

\bibitem{interlayer_SWJang2019} Jang S W, Jeong M Y, Yoon H, Ryee S and Han M J 2019 Microscopic understanding of magnetic interactions in bilayer CrI$_3$ \textit{Phys. Rev. Mater.} \textbf{3} 031001(R)

\bibitem{CrI3_Song2019_Pressure} Tiancheng Song, Zaiyao Fei, Matthew Yankowitz, Zhong Lin, Qianni Jiang, Kyle Hwangbo, Qi Zhang, Bosong Sun, Takashi Taniguchi, Kenji Watanabe, Michael A. McGuire, David Graf, Ting Cao, Jiun-Haw Chu, David H. Cobden, Cory R. Dean, Di Xiao and Xiaodong Xu, Switching 2D magnetic states via pressure tuning of layer stacking \textit{Nat. Mat.} \textbf{18} 1298–1302 (2019).

\bibitem{CrI3_Li2019_Pressure} Li T \textit{et al} 2019 Pressure-controlled interlayer magnetism in atomically thin CrI$_3$ \textit{Nat. Mat.} \textbf{18} 1303-8

\bibitem{CrBr3_Zhang2019} Zhang Z \textit{et al} 2019 Direct Photoluminescence Probing of Ferromagnetism in Monolayer Two-Dimensional CrBr$_3$ \textit{Nano Lett.} \textbf{19} 3138–42

\bibitem{CrCl3_Klein2019} Klein D R \textit{et al} 2019 Enhancement of interlayer exchange in an ultrathin two-dimensional magnet \textit{Nat. Phys.} \textbf{15}, 1255-60

\bibitem{CrI3_Zhong2017} Zhong D \textit{et al} 2017 Van der Waals engineering of ferromagnetic semiconductor heterostructures for spin and valleytronics \textit{Sci. Adv.} \textbf{3} e1603113

\bibitem{CrBr3_Ghazaryan2018} Ghazaryan D \textit{et al} 2018 Magnon-assisted tunnelling in van der Waals heterostructures based on CrBr$_3$ \textit{Nat. Elect.} \textbf{1} 344–9 

\bibitem{DiracMaterials2014} Wehling T O, Black-Schaffer A M and Balatsky A V 2014 Dirac Materials \textit{Advances in Physics} \textbf{63} 1-76

\bibitem{DiracMagnon2018} Pershoguba S S \textit{et al} 2018 \textit{Phys. Rev. X} \textbf{8} 011010

\bibitem{graphene} Geim A K and Novoselov K S 2007 The rise of graphene \textit{Nat. Mater.} \textbf{6} 183–91

\bibitem{CrBr3_Samuelson1971}  Samuelsen E J, Silberglitt R, Shirane G and Remeika J P 1971 Spin Waves in Ferromagnetic CrBr$_3$ Studied by Inelastic Neutron Scattering \textit{Phys. Rev. B} {\bf 3} 157-66

\bibitem{CoTiO3_Yuan2020} Yuan B \textit{et al} 2020 Dirac Magnons in a Honeycomb Lattice Quantum XY Magnet CoTiO$_3$ \textit{Phys. Rev. X} \textbf{10} 011062

\bibitem{CoTiO3_Elliot2021} Elliot M \textit{et al} Order-by-disorder from bond-dependent exchange and intensity signature of nodal quasiparticles in a honeycomb cobaltate \textit{Nat. Comm.} \textbf{12} 3936

\bibitem{Cu3TeO6_Yao2018} Yao W \textit{et al} 2018 Topological spin excitations in a three-dimensional antiferromagnet \textit{Nat. Phys.} \textbf{14} 1011–5

\bibitem{Cu3TeO6_Bao2018} Bao S \textit{et al} 2018 Discovery of coexisting Dirac and triply degenerate magnons in a three-dimensional antiferromagnet \textit{Nat. Comm.} \textbf{9} 2591

\bibitem{Haldane1988} Haldane F D M 1988 Model for a Quantum Hall Effect without Landau Levels: Condensed-Matter Realization of the "Parity Anomaly" \textit{Phys. Rev. Lett.} \textbf{61} 2015-8

\bibitem{kanemele2005} Kane C L and Mele E J 2005 Quantum Spin Hall Effect in Graphene \textit{Phys. Rev. Lett.} \textbf{95} 226801

\bibitem{Owerre2016} Owerre S A 2016 A first theoretical realization of honeycomb topological magnon insulator \textit{J. Phys. Condens. Matter} \textbf{28} 386001

\bibitem{SKKim2016} Kim S K, Ochoa H, Zarzuela R and Tserkovnyak Y 2016 Realization of the Haldane-Kane-Mele Model in a System of Localized Spins \textit{Phys. Rev. Lett.} \textbf{117} 227201

\bibitem{LBChen_PRB2020} Chen L \textit{et al} 2020 Magnetic anisotropy in ferromagnetic CrI$_3$ \textit{Phys. Rev. B.} \textbf{101} 134418

\bibitem{CrBr3_Cai2021} Cai Z \textit{et al} 2021 Topological magnon insulator spin excitations in the two-dimensional ferromagnet CrBr$_3$ \textit{Phys. Rev. B} \textbf{104} L020402

\bibitem{CrCl3_McGuire2017} McGuire M A \textit{et al} 2017 \textit{Phys. Rev. Mater.} \textbf{1} 014001

\bibitem{CrCl3_Dupont2021} M. Dupont, Y. O. Kvashnin, M. Shiranzaei, J. Fransson, N. Laflorencie, and A. Kantian, Phys. Rev. Lett. \textbf{127}, 037204 (2021).

\bibitem{CrCl3_Lu2020} Lu X, Fei R, Zhu L and Yang L 2020 Meron-like topological spin defects in monolayer CrCl$_3$ \textit{Nat. Comm.} \textbf{11} 4724

\bibitem{CrCl3_Seri2020} Serri M \textit{et al} 2020 Enhancement of the Magnetic Coupling in Exfoliated CrCl$_3$ Crystals Observed by Low-Temperature Magnetic Force Microscopy and X-ray Magnetic Circular Dichroism \textit{Adv. Mater.} \textbf{32} 2000566

\bibitem{SEQUOIA} Granroth G E \textit{et al} 2010 \textit{J. Phys. Conf. Ser.} \textbf{251} 012058

\bibitem{HYSPEC} Winn B \textit{et al} 2015 \textit{EPJ Web Conf.} \textbf{83} 03017

\bibitem{SpinW} Toth S and Lake B 2015 Linear spin wave theory for single-Q incommensurate magnetic structures \textit{J. Phys.: Condens. Matter} \textbf{27} 166002

\bibitem{CrI3_McGuire2014} McGuire M A, Dixit H, Cooper V R and Sales B C 2015 Coupling of Crystal Structure and Magnetism in the Layered, Ferromagnetic Insulator CrI$_3$ \textit{Chem. Mater.} \textbf{27} 612-20

\bibitem{SI} See the Supplemental Information for additional data.

\bibitem{CrI3_Niu2020} Niu B \textit{et al} 2020 Coexistence of Magnetic Orders in Two-Dimensional Magnet CrI$_3$ \textit{Nano Lett.} \textbf{20} 553-8

\bibitem{CrCl3_MacNeill2019} MacNeill D \textit{et al} 2019 Gigahertz Frequency Antiferromagnetic Resonance and Strong Magnon-Magnon Coupling in the Layered Crystal CrCl$_3$ \textit{Phys. Rev. Lett.} \textbf{123} 047204

\bibitem{mmi1} Loly P D and Doniach S 1968 Estimate of Damping of Short-Wave Magnons for a Heisenberg Ferromagnet below $T_\mathrm{C}$ \textit{Phys. Rev.} \textbf{173} 603-4

\bibitem{mmi2} Wang K \textit{et al} 2020 Magnon-magnon interaction and magnon relaxation time in a ferromagnetic Cr$_2$Ge$_2$Te$_6$ monolayer \textit{Phys. Rev. B} \textbf{102} 235434

\bibitem{mpc1} Woods L M 2001 Magnon-phonon effects in ferromagnetic manganites \textit{Phys. Rev. B} \textbf{65} 014409

\bibitem{CrI3_Cenker2021} Cenker J \textit{et al} 2021 Direct observation of two-dimensional magnons in atomically thin CrI$_3$ \textit{Nat. Phys.} \textbf{17} 20-5

\end{thebibliography}
\end{document}